\documentclass[%
preprint,
 amsmath,amssymb,
 aps,
]{revtex4-2}

\usepackage{float}
\usepackage{subfig}
\usepackage{graphicx}% Include figure files
\usepackage{dcolumn}% Align table columns on decimal point
\usepackage{bm}% bold math
\usepackage[mathlines]{lineno}% Enable numbering of text and display math

\newcommand{\lirep}[2]{\textcolor{red}{\sout{#1}}\textcolor{blue}{\uline{#2}}}
\renewcommand{\lirep}[2]{#2}      %反注释此行即可接受所有修改

\newcommand{\lidel}[1]{\lirep{#1}{}}
\newcommand{\liadd}[1]{\lirep{}{#1}}

\begin{document}

\title{Optimal gamma-ray selections for monochromatic line searches with DAMPE}% Force line breaks with \\

\author{Zun-Lei Xu$^{1,2}$}
\author{Kai-Kai Duan$^{1}$}
\author{Wei Jiang$^{1}$}
\author{Shi-Jun Lei$^{1}$}
\author{Xiang Li$^{1,2}$\footnote{Corresponding author: xiangli@pmo.ac.cn}}
\author{Zhao-Qiang Shen$^{1}$\footnote{Corresponding author: zqshen@pmo.ac.cn}}
\author{Tao Ma$^{1,2}$}
\author{Meng Su$^1$}
\author{Qiang Yuan$^{1,2}$}
\author{Chuan Yue$^1$}
\author{Yi-Zhong Fan$^{1,2}$}
\author{Jin Chang$^{1,2}$}

\affiliation{
$^1$Key Laboratory of Dark Matter and Space Astronomy, Purple Mountain Observatory, Chinese Academy of Sciences, Nanjing 210023, China\\
$^2$School of Astronomy and Space Science, University of Science and Technology of China, Hefei 230026, China\\
}

\date{\today}% It is always \today, today,
             %  but any date may be explicitly specified

\begin{abstract}
The DArk Matter Particle Explorer (DAMPE) is a space high-energy cosmic-ray detector covering a \lirep{wild}{wide} energy band with a high energy resolution. One of the key scientific goals of DAMPE is to carry out indirect detection of dark matter by searching for high-energy gamma-ray line structure. To promote the sensitivity of gamma-ray line search with DAMPE, it is crucial to improve the acceptance and energy resolution of gamma-ray photons. In this paper, we quantitatively prove that the photon sample with the largest ratio of acceptance to energy resolution is optimal for line search. We therefore develop a line-search sample specifically optimized for the line search. Mean\lidel{ }while, in order to increase the statistics, we also selected the so called BGO-only photons that convert into $e^+e^-$ pairs only in the BGO calorimeter. The standard, the line-search, and the BGO-only photon samples are then tested for line search individually and collectively. The results show that a significantly improved limit could be obtained from an appropriate combination of the date sets, and the increase is about 20\% for the highest case \lirep{comparing}{compared with} using the standard sample only.
\end{abstract}

%\keywords{Suggested keywords}%Use showkeys class option if keyword
                              %display desired
\maketitle

%\tableofcontents YuYH2017

\section{Introduction}
The DArk Matter Particle Explorer (DAMPE) is an advanced high-energy cosmic-ray detector in orbit \cite{ChangJ2014,DmpMission}. The DAMPE consists of four sub detectors \cite{ChangJ2014,DmpMission}. The Plastic Scintillation Detector (PSD) at the top is mainly used for charge measurement \cite{YuYH2017}. The Silicon-Tungsten tracKer converter detector (STK) is mainly used for track measurement and photon conversion. The BGO calorimeter (BGO) is used to measure the energy of incident particles with high precision and to distinguish electrons and protons \cite{ZhangZY2015,ZhangZY2016}. The Neutron Detector (NUD) at the bottom is used to assist in distinguishing electrons and protons \cite{HuangYY2020}. DAMPE has obtained very precise measurements of energy spectra of cosmic ray electrons and positrons, protons, and helium nuclei \cite{DmpElectron,DmpProton,Alemanno2021}, which offers important implications in understanding the physics about cosmic rays \cite{Yue2019,Yuan2020,Yuan2018}.

Gamma-ray observation is one of the three major scientific targets of DAMPE, which can be used in the research of gamma-ray astronomy, time domain astronomy, and indirect detection of dark matter. The gamma-ray line \liadd{above 10~GeV} is one of the most important characteristic signals in the indirect detection of dark matter, as hardly any other astrophysical process can produce such kind of structure.  Many works have been carried out on gamma-ray line search using the Fermi-LAT data, but only some \lirep{suspected}{tentative} signals with low confidence are found so far \cite{Charles2016,Huang2012,Anderson2016,Liang2017,LiS2019,Mazziotta2020}, such as the \lidel{suspected} line \lirep{signal}{candidate} found at about 43 GeV, which can also be interpreted as the product of annihilation of dark matter particles \cite{Liang2016,Shen2021}. Given the high energy resolution of DAMPE among the detectors of similar kind \cite{DmpCalibration}, it has a unique advantage in the search of gamma-ray line signal.

The number of photons in a line signal is proportional to the acceptance of the instrument, while the number of background photons underlaying the line structure is proportional to the product of the acceptance and the energy dispersion width. To improve the \lirep{significance of suspected signal}{sensitivity}, the key issue is to improve the acceptance and energy resolution of the detector. In section 2, We quantitatively prove that the sensitivity of gamma-ray line signal search is positively correlated with the ratio of acceptance to energy resolution. We have developed a standard gamma-ray data sample before \cite{XuZL2018,Duan2019}, which, however, is not optimized for line search. In order to maximize the sensitivity of gamma-ray line search, it is necessary to balance the acceptance and energy resolution in the photon selection algorithm. We therefore developed a photon sample specifically optimized for line search, and this line-search sample is described in detail in section 3. \liadd{On the other hand,} \lirep{T}{t}he increase of photon statistics can significantly improve the sensitivity of line search. For this purpose, we also find in this work the photons that convert into $e^+e^-$ pairs in the BGO calorimeter. With only BGO track but no STK track, these photons of poor angular resolution are but useful for line search. The selection of these BGO-only photons is introduced in section 4. In section 5, we use the standard, the line-search, and the BGO-only data set for line search respectively. The results show that the constraint given by the new data set is significantly stronger than that given by the standard data set, showing that the new data sets are optimized for gamma-ray line search.
% The \nocite command causes all entries in a bibliography to be printed out
% whether or not they are actually referenced in the text. This is appropriate
% for the sample file to show the different styles of references, but authors
% most likely will not want to use it.

\section{Sensitivity for linelike structures}
Acceptance is the integral of the effective area over the solid angle \cite{Acc2011}, and we use the ratio of acceptance $\mathcal{A}$ to the half width of 68\% energy containment $\Delta E/E$ as the optimize target for the LineSearch data set, since this quantity is positively correlated to the signal-to-noise ratio of lines.
Qualitatively, the local significance of a line structure can be written as $n_{\rm line}/\sqrt{n_{\rm bkg,eff}}$, where $n_{\rm line}$ and $n_{\rm bkg,eff}$ represent the photon counts from the line and from the background emission, respectively.
The line counts from given targets are proportional to the effective area $\epsilon$ and the observing time $T$, i.e. $n_{\rm line} \approx F_{\rm line}\times \epsilon(E) T$. Since the diffuse emission from the Milky Way halo is our target and DAMPE mostly stays in the survey mode, the time and space averaged target direction in the detector reference frame is quite uniform. Therefore we have $n_{\rm line} \propto \mathcal{A}$ according to the definition of acceptance.
Concerning the background counts, since only the background fluctuation under the line can directly expose or hide a signal, we use the counts of background emission under the line peak rather than the counts in the entire energy range as $n_{\rm bkg,eff}$, i.e. the ``effective background'' in previous works \cite{Ackermann2013a,Albert2014}. The effective background counts can be approximated using the counts integrated around the line energy $E$, i.e. $n_{\rm bkg,eff} \approx F_{\rm bkg}(E) \epsilon(E) T \cdot 2 \Delta E$, where $F_{\rm bkg}(E)$ is the spectrum of background emission within the region of interest, and $\Delta E$ is the half width of the line.
Using the same argument above, $n_{\rm bkg,eff} \propto \mathcal{A}\times\Delta E/E$.
Therefore the significance of a line improves with the quantity,
\begin{equation}
    \frac{n_{\rm line}}{\sqrt{n_{\rm bkg,eff}}} \propto \frac{\mathcal{A}}{\sqrt{\mathcal{A}\times \Delta E/E}} = \sqrt{\frac{\mathcal{A}}{\Delta E/E}}.
\end{equation}

On the other hand, when no line signal exists, the 95\% confidence level upper limit of counts is $1.64\sigma$ deviated from the null model, so we have $n_{\rm line,UL} = 1.64\sqrt{n_{\rm bkg,eff}}$.
Therefore the upper limit on the line spectrum is
\begin{equation}
    F_{\rm line,UL} = \frac{n_{\rm line,UL}}{\epsilon\, T}\, \delta (E-E_{\rm line}) \propto \frac{\sqrt{\mathcal{A}\times \Delta E/E}}{\mathcal{A}} = \sqrt{(\Delta E/E)/\mathcal{A}}.
\end{equation}

\begin{figure*}[!ht]
  \centering
  \subfloat[]{
    \includegraphics[scale=0.5]{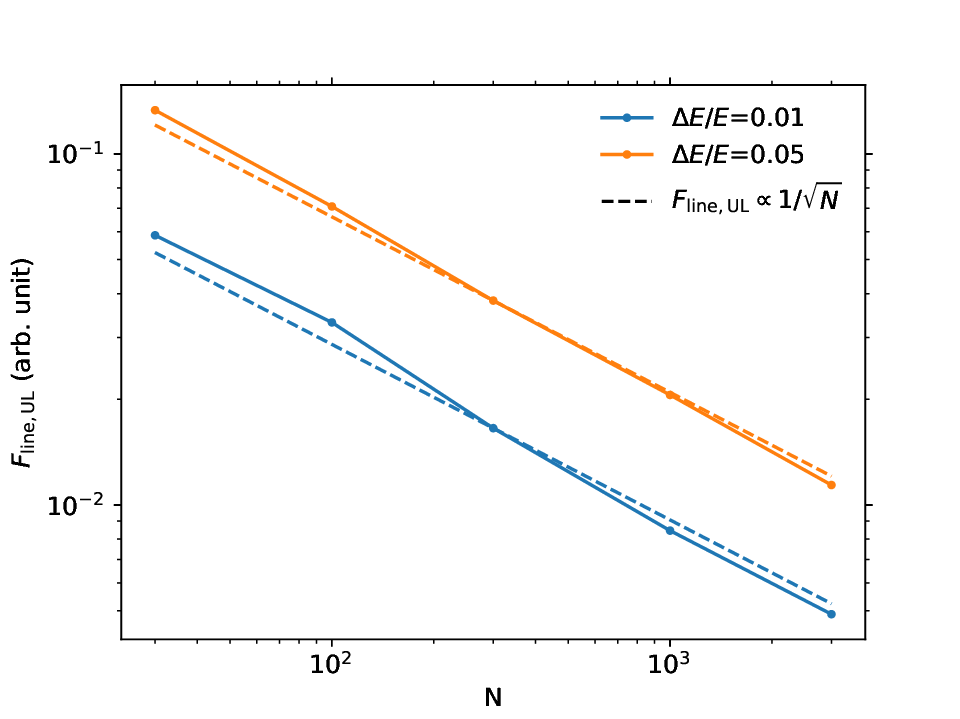}
  }
  \subfloat[]{
    \includegraphics[scale=0.5]{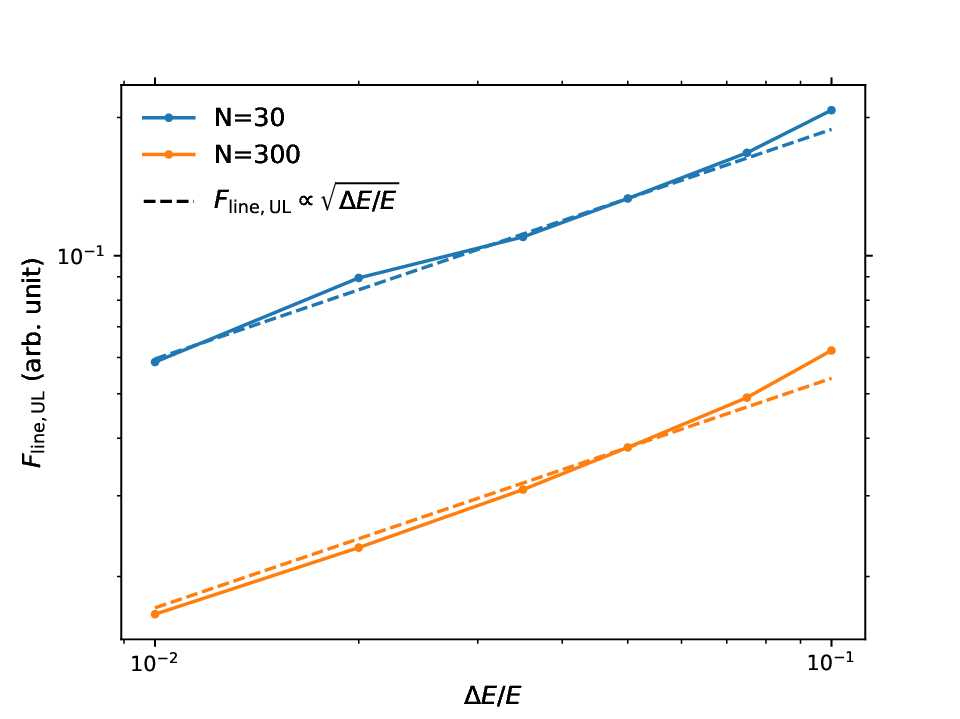}
  }
  \caption{The 95\% confidence level upper limits of line flux for different photon counts (left panel) and energy resolution (right panel).
        The points and solid lines corresponds to the median upper limits in the simulations.
        The dashed lines show the expected relation presented in the paper.
        Please note that at a given energy range and for a particular set of target sources, the photon counts are proportional to the acceptance.}
  \label{Fig1}
\end{figure*}

We also validate the relation with a toy Monte Carlo simulation. Events from 25 GeV to 75 GeV are firstly simulated using a powerlaw background model with spectral index of $-2.5$, then a model consisting of a powerlaw model and a Gaussian line peaked at 50 GeV is adopted to fit the data, and finally the 95\% confidence level upper limits of line flux are calculated.
The photon number in the simulation and the width of the Gaussian line in the fittings are scanned and the results are drawn with points and solid lines in Fig.~\ref{Fig1}.
Generally, the simulation support the relation $F_{\rm line,UL} \propto \sqrt{(\Delta E/E)/\mathcal{A}}$ considering that the photon number is proportional to the acceptance.

\section{line-search sample}

The photons account a very small portion in the cosmic-rays, and we have developed an efficient algorithm to identify the photons before. We estimate that the electrons and protons mixed into the photon sample selected as a background are less than 1\% \cite{XuZL2018}. In the published version, in order to obtain a relatively large photon acceptance and a strong background suppression level, we require the track to pass through the first four layers of BGO calorimeter. According to the DAMPE geometry, that is, to discard the events where the $Z$ direction value of the track is less than 160 mm when the track passes through BGO calorimeter.

Here, we define the variable $R = \frac{\mathcal{A}/{\rm cm^2sr}}{\Delta E/E}$ as the ratio of the acceptance to the energy resolution.
In the last section, we show that a larger $R$ is required to improve the sensitivity of gamma-ray line search. In practice, however, there is a trade-off between these two quantities. A smaller $Z$ value in the BGO through which the particle track is required to pass leads to a larger field of view, which improves the acceptance. On the other hand, a smaller $Z$ value also means that \lidel{more} particles leak \liadd{more} energy outside of the BGO detector, resulting in a poor energy resolution. And vice versa if a larger $Z$ value is required. We thus need to find an optimal $Z$ value to maximize the $R$ value.

\begin{figure}
\centering
\includegraphics[scale=0.7]{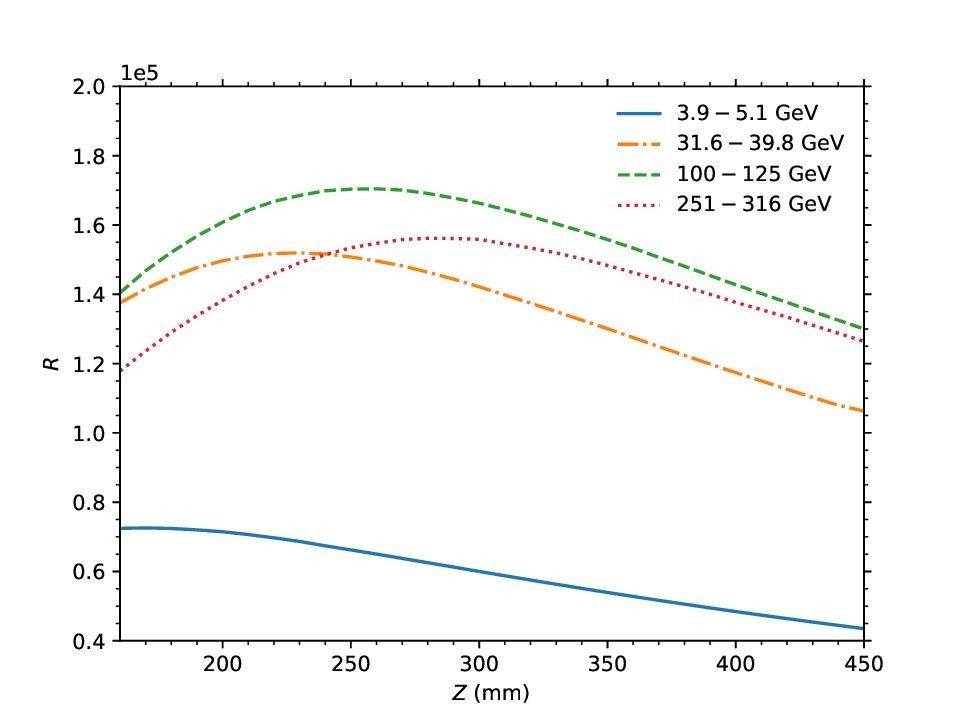}
\caption{The relationship between the $R$ value and the $Z$ value at different energy.}
\label{Fig2}
\end{figure}

\begin{figure}
\centering
\includegraphics[scale=0.7]{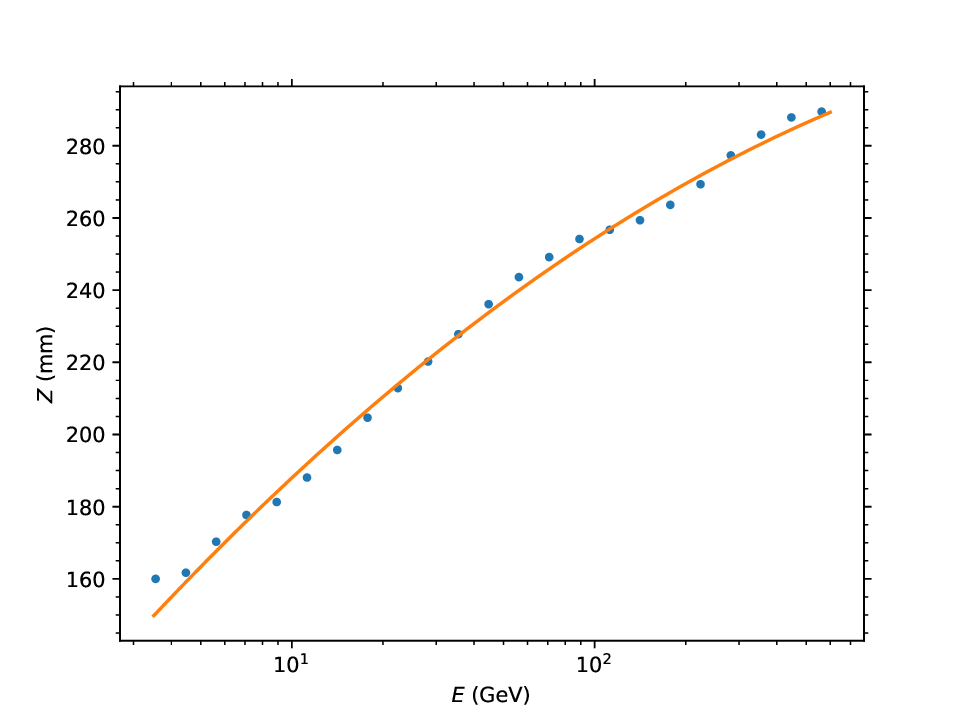}
\caption{The functional relationship between the optimal $Z$ values and different energies.}
\label{Fig3}
\end{figure}

We simulate the relationship between the $R$ value and the $Z$ value at different energ\lirep{y}{ies}. It can be seen from Fig.~\ref{Fig2} that $Z$ takes different value for different $R$, mainly because the acceptance and energy resolution does not change proportionally with $Z$. \lirep{With increasing $Z$}{As $Z$ increases}, the acceptance becomes smaller, \lirep{but for}{while} the energy resolution \lirep{, the improvement is not so obvious}{improves fast first and then relatively stable} after reaching a certain level. This effect is more significant at low energy. It can also be seen from Fig.~\ref{Fig2} that a higher energy leads to a larger $Z$ value when $R$ reaches the maximum. As shown in Fig.~\ref{Fig3}, we use the polynomial \lirep{$Z=97.78+102.17 \times E-11.96 \times E^2$}{$\frac{Z}{\mathrm{mm}}=97.78+102.17 \times \frac{\log E}{\mathrm{GeV}}-11.96 \times \left(\frac{\log E}{\mathrm{GeV}}\right)^2$} to fit the $Z$ value corresponding to the maximum $R$ in different energy, and we can get the \lirep{functional}{analytical} relationship between the optimal $Z$ values and different energ\lirep{y}{ies}.

\begin{figure}
\centering
\includegraphics[scale=0.7]{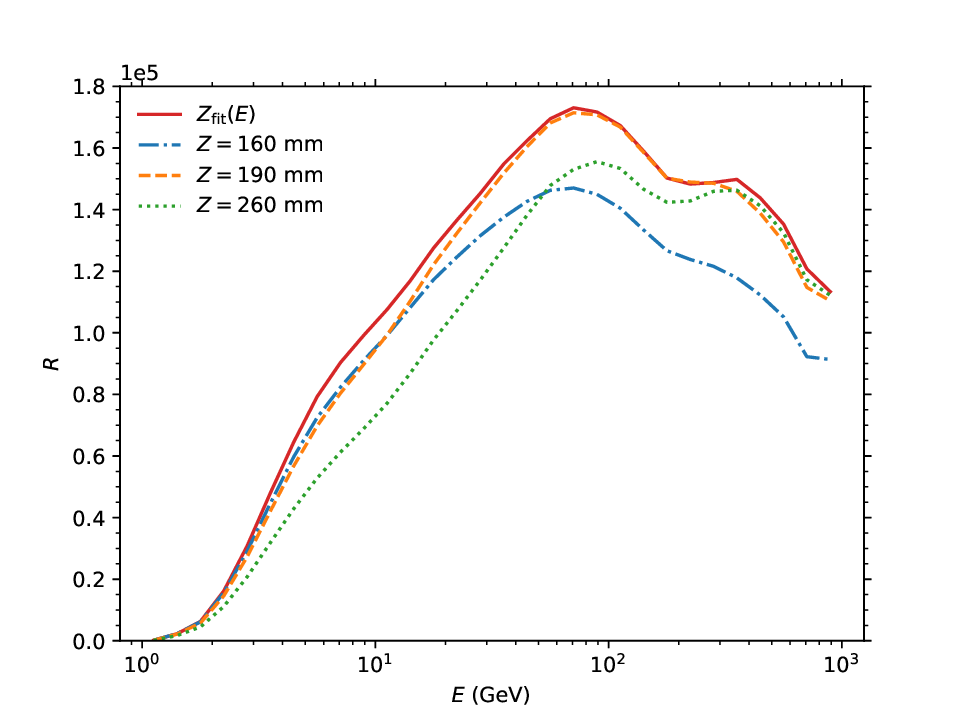}
\caption{The $R$ value at different $Z$ values and energies.}
\label{Fig4}
\end{figure}

As shown in Fig.~\ref{Fig4}, we use the fitted optimal $Z$ value to get the $R$ curves of different energies. We can see that the $R$ value obtained by this method is better than a fixed $Z$ value. \lidel{In this paper, we will use the gamma-ray events obtained by this method to search the line.}

\section{BGO-only sample}

About 10\% of the photons have electron pair conversion in the PSD and other supporting materials before entering the STK. The tungsten plate in the STK is 3 mm in total, which is equivalent to about one radiation length. And only about 50\% of the photons have electron pair conversion in the STK, the rest about 40\% of the photons have electron pair conversion in the BGO. With only BGO track but no STK track, these photons \lidel{only} have poor\liadd{er} direction measurements. The search of line requires as many photons as possible with high energy resolution, and the angular resolution is not a critical issue here, so it is necessary to count in the photons with only BGO track.

\begin{figure*}[!ht]
  \centering
  \subfloat[]{
    \includegraphics[scale=0.5]{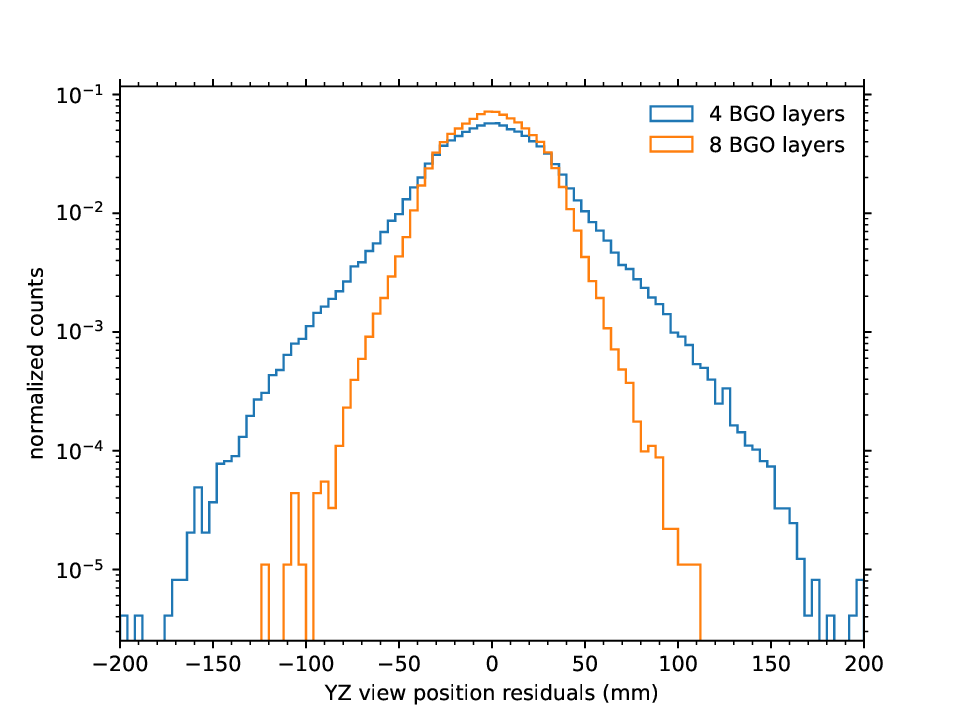}
  }
  \subfloat[]{
    \includegraphics[scale=0.5]{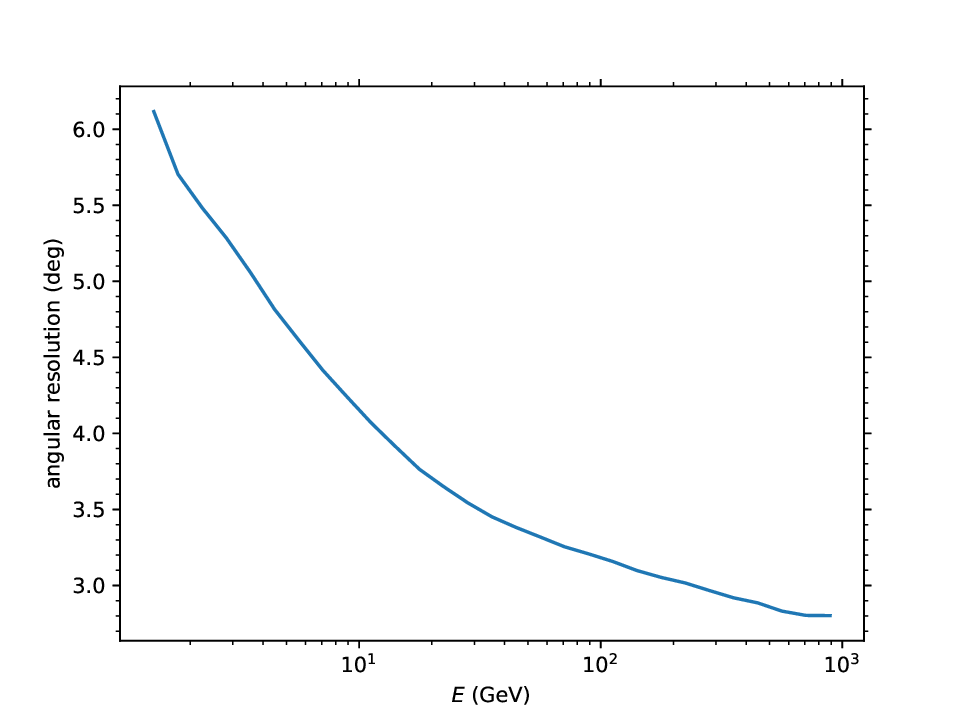}
  }
  \caption{The position resolution of BGO track of 10 GeV electron in the PSD (left) and angular resolution of the BGO track (right).}
  \label{Fig5}
\end{figure*}

The selection process of the photons with only BGO track is similar to that of the photons with STK track. Firstly, most protons are rejected according to the shower morphology of the incident particles in the BGO calorimeter, and then the BGO track is reconstructed. Finally, the PSD crystal through which the track passes is identified to determine whether the crystal has signal to partition photons and charged particles. \liadd{The challenge is that} \lirep{I}{i}f the reconstructed track deviates greatly from the original one, a wrong PSD crystal will be found and a background event maybe introduced. The BGO track reconstruction is based on the centroid method. More accurate centroid determined with more layers result into a track of higher quality. Fig.~\ref{Fig5} shows the position resolution of BGO track of 10 GeV electron in the PSD. The track passing through 8 layers of BGO crystals is obviously better than that passing through only 4 layers. Here we choose the photons passing through more than 8 layers of BGO.

\begin{figure}
\centering
\includegraphics[scale=0.7]{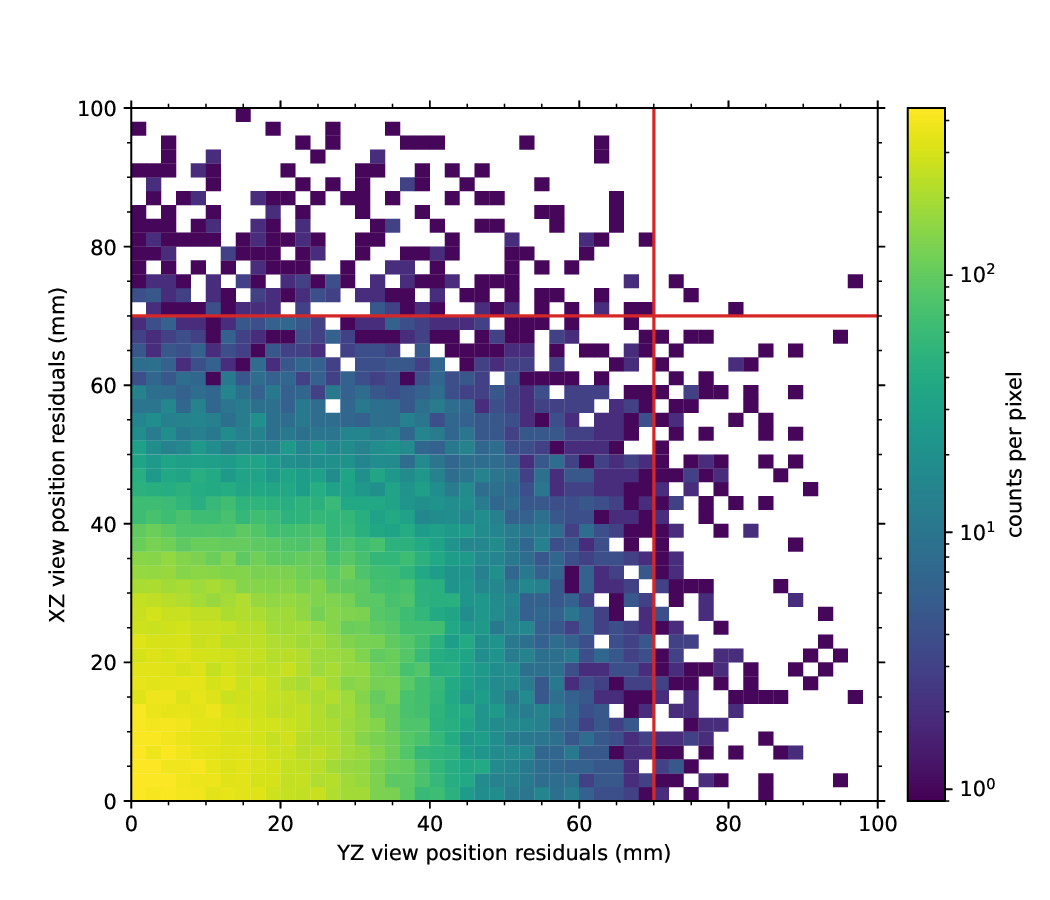}
\caption{The position errors of electron events.}
\label{Fig6}
\end{figure}

The angular resolution of the BGO track is much worse than that of the STK track. As shown in Fig.~\ref{Fig5}, it is less than 3 degrees below 200 GeV. Here the BGO track is reconstructed in the XZ and YZ planes independently. Each PSD crystal is 28 mm wide, and the boundary cut we define at the top is 70 mm. As can be seen from Fig.~\ref{Fig6}, for the 300000 simulated electron events, there are quite a few cases with a position error at the PSD larger than 70 mm in the XZ or YZ plane, but there are \lidel{only a} few events with the position error greater than 70mm in both XZ and YZ planes. As previously mentioned, the PSD has an XY two-layer structure, which can be used for anti coincidence. In this way, when using the BGO track to suppress electrons or protons, it is necessary to jointly judge the two-layer crystal with \lirep{plastic flash}{PSD}. At the same time, it is necessary to search whether there is no signal in the three crystals whose track points to the periphery of the crystal, so as to eliminate the background due to the error in position measurement. We also studied the inhibition ability of the above methods on the electronic background through MC simulation, and the result show that the mixing of electronic background is less than 1\%.

\begin{figure}
\centering
\includegraphics[scale=0.7]{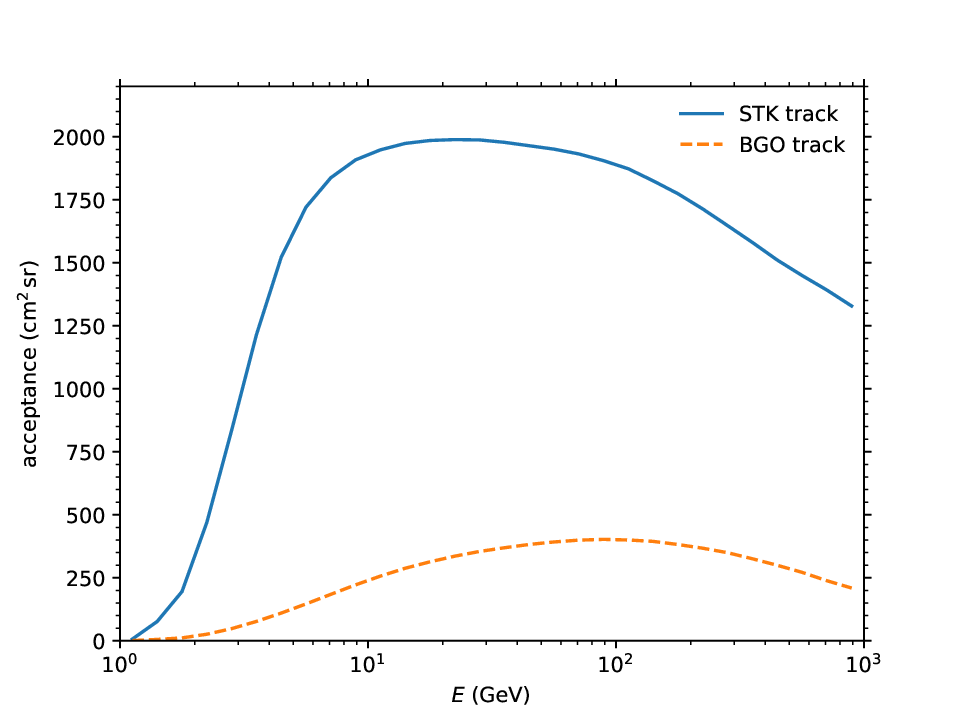}
\caption{The acceptance of photons with only BGO track and the photons with STK track.}
\label{Fig7}
\end{figure}

As shown in Fig.~\ref{Fig7}, at present, the acceptance of photons with only BGO track is only about one tenth of that of the photons with STK track (in the low-energy), which is consistent with the ratio of the number of cases between the two data. The reason of the small acceptance of the photons with only BGO track is that the electron pair conversion of these photons occurs in the BGO, and thus the energy deposited in the first layer of BGO is less, which leads to a low\liadd{er} trigger efficiency \cite{Zhang2019}.

\section{The improvement on the expected line upper limit}
\begin{figure}
\centering
\includegraphics[scale=0.7]{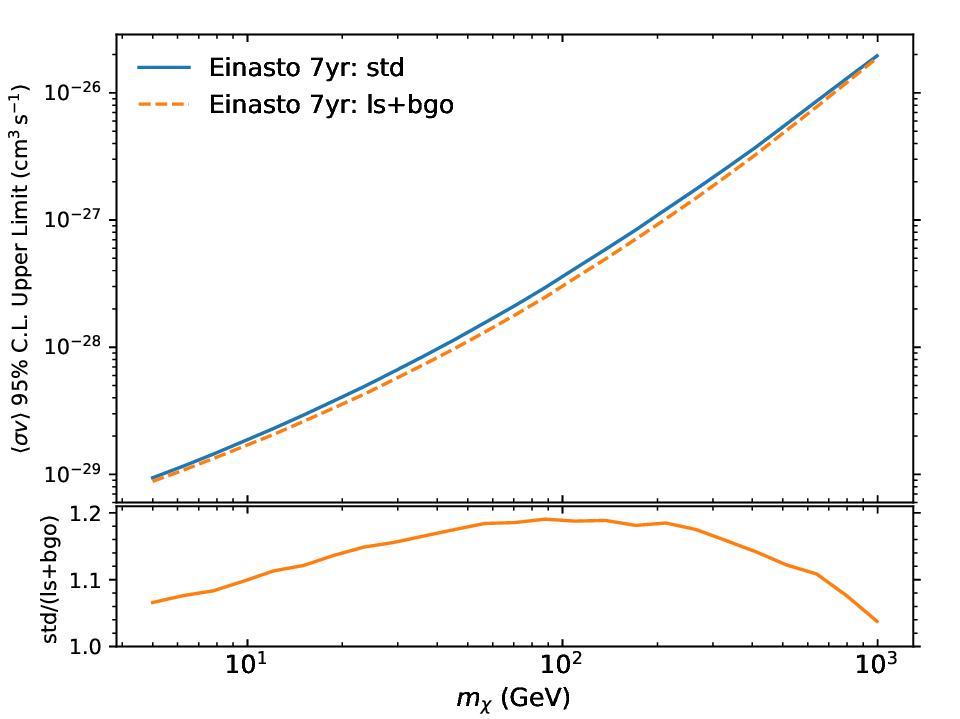}
\caption{Expected 95\% confidence level upper limits of the cross section for dark matter annihilation into a pair of gamma-rays with 7-yr DAMPE data.
        We assume that the dark matter density follows the Einasto profile.
        Blue solid line and orange dashed line represent the upper limit derived using the standard data set and the combination of line-search and BgoOnly data sets, respectively.
        The lower panel shows the ratio of two upper limits..}
\label{Fig8}
\end{figure}

A comparison is made \lirep{when either}{between the using of} the standard data set \lirep{or}{and} the combination of line-search and BGO-only sample\lidel{ is used}. We assume that the density of dark matter follows the Einasto profile with $\alpha=0.17$ \cite{Einasto1965,Navarro2010}. Following, we define the region of interest (ROI) as a $16^\circ$ circle centered on the Galactic center with the Galactic plane region ($|l|>6^\circ$, $|b|<5^\circ$) masked. The background emission is modeled using the average spectrum of the Galactic diffuse model {\tt gll\_iem\_v07.fit}, the isotropic spectrum of {\tt iso\_P8R2\_ULTRACLEAN\_V6\_v01.txt}, and the Fermi-LAT 4FGL point source catalog \cite{FermiGDE2016,4FGL2019} within the ROI. The expected upper limit on the cross section of dark matter annihilation into a pair of photons is calculated and shown in Fig.~\ref{Fig8}. The constraint with the new data sets is stronger than that using the standard data set with the largest improvement being $\sim 20\%$.

\begin{acknowledgements}
The DAMPE mission is funded by the strategic priority science and technology projects in space science of Chinese Academy of Sciences.
In China the data analysis is supported in part by the National Key Research and Development Program of China (No. 2016YFA0400200),
the National Natural Science Foundation of China (Nos. U1738210, U1738123, U1738205, U1738138, 11921003, 12003074), the Youth Innovation Promotion Association CAS, the Key Research Program of the Chinese Academy of Sciences Grant (No. ZDRW-KT-2019-5), and the Entrepreneurship and Innovation Program of Jiangsu Province. In Europe the activities and data analysis are supported by the Swiss National Science Foundation (SNSF), Switzerland, the National Institute for Nuclear Physics (INFN), Italy.
\end{acknowledgements}

\nocite{*}

\bibliography{mybib}% Produces the bibliography via BibTeX.

%apsrev4-2.bst 2019-01-14 (MD) hand-edited version of apsrev4-1.bst
%Control: key (0)
%Control: author (8) initials jnrlst
%Control: editor formatted (1) identically to author
%Control: production of article title (0) allowed
%Control: page (0) single
%Control: year (1) truncated
%Control: production of eprint (0) enabled
\begin{thebibliography}{31}%
\makeatletter
\providecommand \@ifxundefined [1]{%
 \@ifx{#1\undefined}
}%
\providecommand \@ifnum [1]{%
 \ifnum #1\expandafter \@firstoftwo
 \else \expandafter \@secondoftwo
 \fi
}%
\providecommand \@ifx [1]{%
 \ifx #1\expandafter \@firstoftwo
 \else \expandafter \@secondoftwo
 \fi
}%
\providecommand \natexlab [1]{#1}%
\providecommand \enquote  [1]{``#1''}%
\providecommand \bibnamefont  [1]{#1}%
\providecommand \bibfnamefont [1]{#1}%
\providecommand \citenamefont [1]{#1}%
\providecommand \href@noop [0]{\@secondoftwo}%
\providecommand \href [0]{\begingroup \@sanitize@url \@href}%
\providecommand \@href[1]{\@@startlink{#1}\@@href}%
\providecommand \@@href[1]{\endgroup#1\@@endlink}%
\providecommand \@sanitize@url [0]{\catcode `\\12\catcode `\$12\catcode
  `\&12\catcode `\#12\catcode `\^12\catcode `\_12\catcode `\%12\relax}%
\providecommand \@@startlink[1]{}%
\providecommand \@@endlink[0]{}%
\providecommand \url  [0]{\begingroup\@sanitize@url \@url }%
\providecommand \@url [1]{\endgroup\@href {#1}{\urlprefix }}%
\providecommand \urlprefix  [0]{URL }%
\providecommand \Eprint [0]{\href }%
\providecommand \doibase [0]{https://doi.org/}%
\providecommand \selectlanguage [0]{\@gobble}%
\providecommand \bibinfo  [0]{\@secondoftwo}%
\providecommand \bibfield  [0]{\@secondoftwo}%
\providecommand \translation [1]{[#1]}%
\providecommand \BibitemOpen [0]{}%
\providecommand \bibitemStop [0]{}%
\providecommand \bibitemNoStop [0]{.\EOS\space}%
\providecommand \EOS [0]{\spacefactor3000\relax}%
\providecommand \BibitemShut  [1]{\csname bibitem#1\endcsname}%
\let\auto@bib@innerbib\@empty
%</preamble>
\bibitem [{\citenamefont {{Chang}}(2014)}]{ChangJ2014}%
  \BibitemOpen
  \bibfield  {author} {\bibinfo {author} {\bibfnamefont {J.}~\bibnamefont
  {{Chang}}},\ }\bibfield  {title} {\bibinfo {title} {{Dark Matter Particle
  Explorer: The First Chinese Cosmic Ray and Hard $\gamma$-ray Detector in
  Space}},\ }\href@noop {} {\bibfield  {journal} {\bibinfo  {journal} {Chinese
  Journal of Space Science}\ }\textbf {\bibinfo {volume} {34}},\ \bibinfo
  {pages} {550} (\bibinfo {year} {2014})}\BibitemShut {NoStop}%
\bibitem [{\citenamefont {{Chang}}\ \emph {et~al.}(2017)\citenamefont {{Chang}}
  \emph {et~al.}}]{DmpMission}%
  \BibitemOpen
  \bibfield  {author} {\bibinfo {author} {\bibfnamefont {J.}~\bibnamefont
  {{Chang}}} \emph {et~al.} (\bibinfo {collaboration} {DAMPE Collaborabtion}),\
  }\bibfield  {title} {\bibinfo {title} {{The DArk Matter Particle Explorer
  mission}},\ }\href {https://doi.org/10.1016/j.astropartphys.2017.08.005}
  {\bibfield  {journal} {\bibinfo  {journal} {Astropart. Phys.}\ }\textbf
  {\bibinfo {volume} {95}},\ \bibinfo {pages} {6} (\bibinfo {year}
  {2017})}\BibitemShut {NoStop}%
\bibitem [{\citenamefont {Yu}\ \emph {et~al.}(2017)\citenamefont {Yu} \emph
  {et~al.}}]{YuYH2017}%
  \BibitemOpen
  \bibfield  {author} {\bibinfo {author} {\bibfnamefont {Y.}~\bibnamefont {Yu}}
  \emph {et~al.},\ }\bibfield  {title} {\bibinfo {title} {{The plastic
  scintillator detector for DAMPE}},\ }\href
  {https://doi.org/10.1016/j.astropartphys.2017.06.004} {\bibfield  {journal}
  {\bibinfo  {journal} {Astropart. Phys.}\ }\textbf {\bibinfo {volume} {94}},\
  \bibinfo {pages} {1} (\bibinfo {year} {2017})},\ \Eprint
  {https://arxiv.org/abs/1703.00098} {arXiv:1703.00098 [astro-ph.IM]}
  \BibitemShut {NoStop}%
\bibitem [{\citenamefont {Zhang}\ \emph {et~al.}(2015)\citenamefont {Zhang}
  \emph {et~al.}}]{ZhangZY2015}%
  \BibitemOpen
  \bibfield  {author} {\bibinfo {author} {\bibfnamefont {Z.}~\bibnamefont
  {Zhang}} \emph {et~al.},\ }\bibfield  {title} {\bibinfo {title} {{Design of a
  high dynamic range photomultiplier base board for the BGO ECAL of DAMPE}},\
  }\href {https://doi.org/10.1016/j.nima.2015.01.036} {\bibfield  {journal}
  {\bibinfo  {journal} {Nucl. Instrum. Meth. A}\ }\textbf {\bibinfo {volume}
  {780}},\ \bibinfo {pages} {21} (\bibinfo {year} {2015})}\BibitemShut
  {NoStop}%
%%CITATION = NUIMA,A780,21;%%
\bibitem [{\citenamefont {{Zhang}}\ \emph {et~al.}(2016)\citenamefont {{Zhang}}
  \emph {et~al.}}]{ZhangZY2016}%
  \BibitemOpen
  \bibfield  {author} {\bibinfo {author} {\bibfnamefont {Z.}~\bibnamefont
  {{Zhang}}} \emph {et~al.},\ }\bibfield  {title} {\bibinfo {title} {{The
  calibration and electron energy reconstruction of the BGO ECAL of the DAMPE
  detector}},\ }\href {https://doi.org/10.1016/j.nima.2016.08.015} {\bibfield
  {journal} {\bibinfo  {journal} {Nucl. Instrum. Meth. A}\ }\textbf {\bibinfo
  {volume} {836}},\ \bibinfo {pages} {98} (\bibinfo {year} {2016})}\BibitemShut
  {NoStop}%
\bibitem [{\citenamefont {{Huang}}\ \emph {et~al.}(2020)\citenamefont
  {{Huang}}, \citenamefont {{Ma}}, \citenamefont {{Yue}}, \citenamefont
  {{Zhang}}, \citenamefont {{Chang}}, \citenamefont {{Dong}},\ and\
  \citenamefont {{Zhang}}}]{HuangYY2020}%
  \BibitemOpen
  \bibfield  {author} {\bibinfo {author} {\bibfnamefont {Y.-Y.}\ \bibnamefont
  {{Huang}}}, \bibinfo {author} {\bibfnamefont {T.}~\bibnamefont {{Ma}}},
  \bibinfo {author} {\bibfnamefont {C.}~\bibnamefont {{Yue}}}, \bibinfo
  {author} {\bibfnamefont {Y.}~\bibnamefont {{Zhang}}}, \bibinfo {author}
  {\bibfnamefont {J.}~\bibnamefont {{Chang}}}, \bibinfo {author} {\bibfnamefont
  {T.-K.}\ \bibnamefont {{Dong}}},\ and\ \bibinfo {author} {\bibfnamefont
  {Y.-Q.}\ \bibnamefont {{Zhang}}},\ }\bibfield  {title} {\bibinfo {title}
  {{Calibration and performance of the neutron detector onboard of the DAMPE
  mission}},\ }\href@noop {} {\bibfield  {journal} {\bibinfo  {journal} {Res.
  Astron. Astrophys.}\ } (\bibinfo {year} {2020})},\ \Eprint
  {https://arxiv.org/abs/2005.07828} {arXiv:2005.07828} \BibitemShut {NoStop}%
\bibitem [{\citenamefont {Ambrosi}\ \emph {et~al.}(2017)\citenamefont {Ambrosi}
  \emph {et~al.}}]{DmpElectron}%
  \BibitemOpen
  \bibfield  {author} {\bibinfo {author} {\bibfnamefont {G.}~\bibnamefont
  {Ambrosi}} \emph {et~al.} (\bibinfo {collaboration} {DAMPE Collaborabtion}),\
  }\bibfield  {title} {\bibinfo {title} {{Direct detection of a break in the
  teraelectronvolt cosmic-ray spectrum of electrons and positrons}},\ }\href
  {https://doi.org/10.1038/nature24475} {\bibfield  {journal} {\bibinfo
  {journal} {Nature}\ }\textbf {\bibinfo {volume} {552}},\ \bibinfo {pages}
  {63} (\bibinfo {year} {2017})},\ \Eprint {https://arxiv.org/abs/1711.10981}
  {arXiv:1711.10981 [astro-ph.HE]} \BibitemShut {NoStop}%
%%CITATION = ARXIV:1711.10981;%%
\bibitem [{\citenamefont {An}\ \emph {et~al.}(2019)\citenamefont {An} \emph
  {et~al.}}]{DmpProton}%
  \BibitemOpen
  \bibfield  {author} {\bibinfo {author} {\bibfnamefont {Q.}~\bibnamefont {An}}
  \emph {et~al.} (\bibinfo {collaboration} {DAMPE Collaborabtion}),\ }\bibfield
   {title} {\bibinfo {title} {{Measurement of the cosmic-ray proton spectrum
  from 40 GeV to 100 TeV with the DAMPE satellite}},\ }\href
  {https://doi.org/10.1126/sciadv.aax3793} {\bibfield  {journal} {\bibinfo
  {journal} {Sci. Adv.}\ }\textbf {\bibinfo {volume} {5}},\ \bibinfo {pages}
  {eaax3793} (\bibinfo {year} {2019})},\ \Eprint
  {https://arxiv.org/abs/1909.12860} {arXiv:1909.12860 [astro-ph.HE]}
  \BibitemShut {NoStop}%
%%CITATION = ARXIV:1909.12860;%%
\bibitem [{\citenamefont {Alemanno}\ \emph {et~al.}(2021)\citenamefont
  {Alemanno} \emph {et~al.}}]{Alemanno2021}%
  \BibitemOpen
  \bibfield  {author} {\bibinfo {author} {\bibfnamefont {F.}~\bibnamefont
  {Alemanno}} \emph {et~al.} (\bibinfo {collaboration} {DAMPE
  Collaborabtion}),\ }\bibfield  {title} {\bibinfo {title} {{Measurement of the
  cosmic ray helium energy spectrum from 70 GeV to 80 TeV with the DAMPE space
  mission}},\ }\href {https://doi.org/10.1103/PhysRevLett.126.201102}
  {\bibfield  {journal} {\bibinfo  {journal} {Phys. Rev. Lett.}\ }\textbf
  {\bibinfo {volume} {126}},\ \bibinfo {pages} {201102} (\bibinfo {year}
  {2021})},\ \Eprint {https://arxiv.org/abs/2105.09073} {arXiv:2105.09073
  [astro-ph.HE]} \BibitemShut {NoStop}%
\bibitem [{\citenamefont {Yue}\ \emph {et~al.}(2020)\citenamefont {Yue} \emph
  {et~al.}}]{Yue2019}%
  \BibitemOpen
  \bibfield  {author} {\bibinfo {author} {\bibfnamefont {C.}~\bibnamefont
  {Yue}} \emph {et~al.},\ }\bibfield  {title} {\bibinfo {title} {{Implications
  on the origin of cosmic rays in light of 10 TV spectral softenings}},\ }\href
  {https://doi.org/10.1007/s11467-019-0946-8} {\bibfield  {journal} {\bibinfo
  {journal} {Front. Phys. (Beijing)}\ }\textbf {\bibinfo {volume} {15}},\
  \bibinfo {pages} {24601} (\bibinfo {year} {2020})},\ \Eprint
  {https://arxiv.org/abs/1909.12857} {arXiv:1909.12857 [astro-ph.HE]}
  \BibitemShut {NoStop}%
\bibitem [{\citenamefont {Yuan}\ \emph {et~al.}(2021)\citenamefont {Yuan},
  \citenamefont {Qiao}, \citenamefont {Guo}, \citenamefont {Fan},\ and\
  \citenamefont {Bi}}]{Yuan2020}%
  \BibitemOpen
  \bibfield  {author} {\bibinfo {author} {\bibfnamefont {Q.}~\bibnamefont
  {Yuan}}, \bibinfo {author} {\bibfnamefont {B.-Q.}\ \bibnamefont {Qiao}},
  \bibinfo {author} {\bibfnamefont {Y.-Q.}\ \bibnamefont {Guo}}, \bibinfo
  {author} {\bibfnamefont {Y.-Z.}\ \bibnamefont {Fan}},\ and\ \bibinfo {author}
  {\bibfnamefont {X.-J.}\ \bibnamefont {Bi}},\ }\bibfield  {title} {\bibinfo
  {title} {{Nearby source interpretation of differences among light and medium
  composition spectra in cosmic rays}},\ }\href
  {https://doi.org/10.1007/s11467-020-0990-4} {\bibfield  {journal} {\bibinfo
  {journal} {Front. Phys. (Beijing)}\ }\textbf {\bibinfo {volume} {16}},\
  \bibinfo {pages} {24501} (\bibinfo {year} {2021})},\ \Eprint
  {https://arxiv.org/abs/2007.01768} {arXiv:2007.01768 [astro-ph.HE]}
  \BibitemShut {NoStop}%
\bibitem [{\citenamefont {Yuan}\ and\ \citenamefont {Feng}(2018)}]{Yuan2018}%
  \BibitemOpen
  \bibfield  {author} {\bibinfo {author} {\bibfnamefont {Q.}~\bibnamefont
  {Yuan}}\ and\ \bibinfo {author} {\bibfnamefont {L.}~\bibnamefont {Feng}},\
  }\bibfield  {title} {\bibinfo {title} {{Dark Matter Particle Explorer
  observations of high-energy cosmic ray electrons plus positrons and their
  physical implications}},\ }\href {https://doi.org/10.1007/s11433-018-9226-y}
  {\bibfield  {journal} {\bibinfo  {journal} {Sci. China Phys. Mech. Astron.}\
  }\textbf {\bibinfo {volume} {61}},\ \bibinfo {pages} {101002} (\bibinfo
  {year} {2018})},\ \Eprint {https://arxiv.org/abs/1807.11638}
  {arXiv:1807.11638 [astro-ph.HE]} \BibitemShut {NoStop}%
\bibitem [{\citenamefont {{Charles}}\ \emph {et~al.}(2016)\citenamefont
  {{Charles}}, \citenamefont {{S{\'a}nchez-Conde}}, \citenamefont {{Anderson}},
  \citenamefont {{Caputo}} \emph {et~al.}}]{Charles2016}%
  \BibitemOpen
  \bibfield  {author} {\bibinfo {author} {\bibfnamefont {E.}~\bibnamefont
  {{Charles}}}, \bibinfo {author} {\bibfnamefont {M.}~\bibnamefont
  {{S{\'a}nchez-Conde}}}, \bibinfo {author} {\bibfnamefont {B.}~\bibnamefont
  {{Anderson}}}, \bibinfo {author} {\bibfnamefont {R.}~\bibnamefont
  {{Caputo}}}, \emph {et~al.},\ }\bibfield  {title} {\bibinfo {title}
  {{Sensitivity projections for dark matter searches with the Fermi large area
  telescope}},\ }\href {https://doi.org/10.1016/j.physrep.2016.05.001}
  {\bibfield  {journal} {\bibinfo  {journal} {Phys. Rep.}\ }\textbf {\bibinfo
  {volume} {636}},\ \bibinfo {pages} {1} (\bibinfo {year} {2016})},\ \Eprint
  {https://arxiv.org/abs/1605.02016} {arXiv:1605.02016} \BibitemShut {NoStop}%
\bibitem [{\citenamefont {{Huang}}\ \emph {et~al.}(2012)\citenamefont
  {{Huang}}, \citenamefont {{Yuan}}, \citenamefont {{Yin}}, \citenamefont
  {{Bi}},\ and\ \citenamefont {{Chen}}}]{Huang2012}%
  \BibitemOpen
  \bibfield  {author} {\bibinfo {author} {\bibfnamefont {X.}~\bibnamefont
  {{Huang}}}, \bibinfo {author} {\bibfnamefont {Q.}~\bibnamefont {{Yuan}}},
  \bibinfo {author} {\bibfnamefont {P.-F.}\ \bibnamefont {{Yin}}}, \bibinfo
  {author} {\bibfnamefont {X.-J.}\ \bibnamefont {{Bi}}},\ and\ \bibinfo
  {author} {\bibfnamefont {X.}~\bibnamefont {{Chen}}},\ }\bibfield  {title}
  {\bibinfo {title} {{Constraints on the dark matter annihilation scenario of
  Fermi 130 GeV gamma-ray line emission by continuous gamma-rays, Milky Way
  halo, galaxy clusters and dwarf galaxies observations}},\ }\href
  {https://doi.org/10.1088/1475-7516/2012/11/048} {\bibfield  {journal}
  {\bibinfo  {journal} {J. Cosmol. Astropart. Phys.}\ }\textbf {\bibinfo
  {volume} {11}},\ \bibinfo {eid} {048}},\ \Eprint
  {https://arxiv.org/abs/1208.0267} {arXiv:1208.0267} \BibitemShut {NoStop}%
\bibitem [{\citenamefont {{Anderson}}\ \emph {et~al.}(2016)\citenamefont
  {{Anderson}}, \citenamefont {{Zimmer}}, \citenamefont {{Conrad}},
  \citenamefont {{Gustafsson}}, \citenamefont {{S{\'a}nchez-Conde}},\ and\
  \citenamefont {{Caputo}}}]{Anderson2016}%
  \BibitemOpen
  \bibfield  {author} {\bibinfo {author} {\bibfnamefont {B.}~\bibnamefont
  {{Anderson}}}, \bibinfo {author} {\bibfnamefont {S.}~\bibnamefont
  {{Zimmer}}}, \bibinfo {author} {\bibfnamefont {J.}~\bibnamefont {{Conrad}}},
  \bibinfo {author} {\bibfnamefont {M.}~\bibnamefont {{Gustafsson}}}, \bibinfo
  {author} {\bibfnamefont {M.}~\bibnamefont {{S{\'a}nchez-Conde}}},\ and\
  \bibinfo {author} {\bibfnamefont {R.}~\bibnamefont {{Caputo}}},\ }\bibfield
  {title} {\bibinfo {title} {{Search for gamma-ray lines towards galaxy
  clusters with the Fermi-LAT}},\ }\href
  {https://doi.org/10.1088/1475-7516/2016/02/026} {\bibfield  {journal}
  {\bibinfo  {journal} {J. Cosmol. Astropart. Phys.}\ }\textbf {\bibinfo
  {volume} {02}}\bibfield  {number} {\bibinfo  {number} { (2)},\ \bibinfo {eid}
  {026}},\ }\Eprint {https://arxiv.org/abs/1511.00014} {arXiv:1511.00014}
  \BibitemShut {NoStop}%
\bibitem [{\citenamefont {{Liang}}\ \emph {et~al.}(2017)\citenamefont
  {{Liang}}, \citenamefont {{Xia}}, \citenamefont {{Duan}}, \citenamefont
  {{Shen}}, \citenamefont {{Li}},\ and\ \citenamefont {{Fan}}}]{Liang2017}%
  \BibitemOpen
  \bibfield  {author} {\bibinfo {author} {\bibfnamefont {Y.-F.}\ \bibnamefont
  {{Liang}}}, \bibinfo {author} {\bibfnamefont {Z.-Q.}\ \bibnamefont {{Xia}}},
  \bibinfo {author} {\bibfnamefont {K.-K.}\ \bibnamefont {{Duan}}}, \bibinfo
  {author} {\bibfnamefont {Z.-Q.}\ \bibnamefont {{Shen}}}, \bibinfo {author}
  {\bibfnamefont {X.}~\bibnamefont {{Li}}},\ and\ \bibinfo {author}
  {\bibfnamefont {Y.-Z.}\ \bibnamefont {{Fan}}},\ }\bibfield  {title} {\bibinfo
  {title} {{Limits on dark matter annihilation cross sections to gamma-ray
  lines with subhalo distributions in N -body simulations and Fermi LAT
  data}},\ }\href {https://doi.org/10.1103/PhysRevD.95.063531} {\bibfield
  {journal} {\bibinfo  {journal} {Phys. Rev. D}\ }\textbf {\bibinfo {volume}
  {95}},\ \bibinfo {eid} {063531} (\bibinfo {year} {2017})},\ \Eprint
  {https://arxiv.org/abs/1703.07078} {arXiv:1703.07078} \BibitemShut {NoStop}%
\bibitem [{\citenamefont {{Li}}\ \emph {et~al.}(2019)\citenamefont {{Li}},
  \citenamefont {{Xia}}, \citenamefont {{Liang}}, \citenamefont {{Duan}} \emph
  {et~al.}}]{LiS2019}%
  \BibitemOpen
  \bibfield  {author} {\bibinfo {author} {\bibfnamefont {S.}~\bibnamefont
  {{Li}}}, \bibinfo {author} {\bibfnamefont {Z.-Q.}\ \bibnamefont {{Xia}}},
  \bibinfo {author} {\bibfnamefont {Y.-F.}\ \bibnamefont {{Liang}}}, \bibinfo
  {author} {\bibfnamefont {K.-K.}\ \bibnamefont {{Duan}}}, \emph {et~al.},\
  }\bibfield  {title} {\bibinfo {title} {{Search for line-like signals in the
  all-sky Fermi-LAT data}},\ }\href
  {https://doi.org/10.1103/PhysRevD.99.123519} {\bibfield  {journal} {\bibinfo
  {journal} {Phys. Rev. D}\ }\textbf {\bibinfo {volume} {99}},\ \bibinfo {eid}
  {123519} (\bibinfo {year} {2019})}\BibitemShut {NoStop}%
\bibitem [{\citenamefont {{Mazziotta}}\ \emph {et~al.}(2020)\citenamefont
  {{Mazziotta}}, \citenamefont {{Loparco}}, \citenamefont {{Serini}},
  \citenamefont {{Cuoco}} \emph {et~al.}}]{Mazziotta2020}%
  \BibitemOpen
  \bibfield  {author} {\bibinfo {author} {\bibfnamefont {M.~N.}\ \bibnamefont
  {{Mazziotta}}}, \bibinfo {author} {\bibfnamefont {F.}~\bibnamefont
  {{Loparco}}}, \bibinfo {author} {\bibfnamefont {D.}~\bibnamefont {{Serini}}},
  \bibinfo {author} {\bibfnamefont {A.}~\bibnamefont {{Cuoco}}}, \emph
  {et~al.},\ }\bibfield  {title} {\bibinfo {title} {{Search for dark matter
  signatures in the gamma-ray emission towards the Sun with the Fermi Large
  Area Telescope}},\ }\href {https://doi.org/10.1103/PhysRevD.102.022003}
  {\bibfield  {journal} {\bibinfo  {journal} {Phys. Rev. D}\ }\textbf {\bibinfo
  {volume} {102}},\ \bibinfo {eid} {022003} (\bibinfo {year} {2020})},\ \Eprint
  {https://arxiv.org/abs/2006.04114} {arXiv:2006.04114} \BibitemShut {NoStop}%
\bibitem [{\citenamefont {{Liang}}\ \emph {et~al.}(2016)\citenamefont
  {{Liang}}, \citenamefont {{Shen}}, \citenamefont {{Li}}, \citenamefont
  {{Fan}} \emph {et~al.}}]{Liang2016}%
  \BibitemOpen
  \bibfield  {author} {\bibinfo {author} {\bibfnamefont {Y.-F.}\ \bibnamefont
  {{Liang}}}, \bibinfo {author} {\bibfnamefont {Z.-Q.}\ \bibnamefont {{Shen}}},
  \bibinfo {author} {\bibfnamefont {X.}~\bibnamefont {{Li}}}, \bibinfo {author}
  {\bibfnamefont {Y.-Z.}\ \bibnamefont {{Fan}}}, \emph {et~al.},\ }\bibfield
  {title} {\bibinfo {title} {{Search for a gamma-ray line feature from a group
  of nearby galaxy clusters with Fermi LAT Pass 8 data}},\ }\href
  {https://doi.org/10.1103/PhysRevD.93.103525} {\bibfield  {journal} {\bibinfo
  {journal} {Phys. Rev. D}\ }\textbf {\bibinfo {volume} {93}},\ \bibinfo {eid}
  {103525} (\bibinfo {year} {2016})},\ \Eprint
  {https://arxiv.org/abs/1602.06527} {arXiv:1602.06527} \BibitemShut {NoStop}%
\bibitem [{\citenamefont {Shen}\ \emph {et~al.}(2021)\citenamefont {Shen},
  \citenamefont {Xia},\ and\ \citenamefont {Fan}}]{Shen2021}%
  \BibitemOpen
  \bibfield  {author} {\bibinfo {author} {\bibfnamefont {Z.-Q.}\ \bibnamefont
  {Shen}}, \bibinfo {author} {\bibfnamefont {Z.-Q.}\ \bibnamefont {Xia}},\ and\
  \bibinfo {author} {\bibfnamefont {Y.-Z.}\ \bibnamefont {Fan}},\ }\bibfield
  {title} {\bibinfo {title} {{Search for line-like and box-shaped spectral
  features from nearby galaxy clusters with 11.4 years of Fermi LAT data}},\
  }\href {https://doi.org/10.3847/1538-4357/ac19ae} {\bibfield  {journal}
  {\bibinfo  {journal} {Astrophys. J.}\ }\textbf {\bibinfo {volume} {920}},\
  \bibinfo {pages} {1} (\bibinfo {year} {2021})},\ \Eprint
  {https://arxiv.org/abs/2108.00363} {arXiv:2108.00363 [astro-ph.HE]}
  \BibitemShut {NoStop}%
\bibitem [{\citenamefont {Ambrosi}\ \emph {et~al.}(2019)\citenamefont {Ambrosi}
  \emph {et~al.}}]{DmpCalibration}%
  \BibitemOpen
  \bibfield  {author} {\bibinfo {author} {\bibfnamefont {G.}~\bibnamefont
  {Ambrosi}} \emph {et~al.} (\bibinfo {collaboration} {DAMPE Collaborabtion}),\
  }\bibfield  {title} {\bibinfo {title} {{The on-orbit calibration of DArk
  Matter Particle Explorer}},\ }\href
  {https://doi.org/10.1016/j.astropartphys.2018.10.006} {\bibfield  {journal}
  {\bibinfo  {journal} {Astropart. Phys.}\ }\textbf {\bibinfo {volume} {106}},\
  \bibinfo {pages} {18} (\bibinfo {year} {2019})},\ \Eprint
  {https://arxiv.org/abs/1907.02173} {arXiv:1907.02173 [astro-ph.IM]}
  \BibitemShut {NoStop}%
%%CITATION = ARXIV:1907.02173;%%
\bibitem [{\citenamefont {Xu}\ \emph {et~al.}(2018)\citenamefont {Xu} \emph
  {et~al.}}]{XuZL2018}%
  \BibitemOpen
  \bibfield  {author} {\bibinfo {author} {\bibfnamefont {Z.-L.}\ \bibnamefont
  {Xu}} \emph {et~al.},\ }\bibfield  {title} {\bibinfo {title} {{An algorithm
  to resolve $\gamma$-rays from charged cosmic rays with DAMPE}},\ }\href
  {https://doi.org/10.1088/1674-4527/18/3/27} {\bibfield  {journal} {\bibinfo
  {journal} {Res. Astron. Astrophys.}\ }\textbf {\bibinfo {volume} {18}},\
  \bibinfo {pages} {027} (\bibinfo {year} {2018})},\ \Eprint
  {https://arxiv.org/abs/1712.02939} {arXiv:1712.02939 [physics.ins-det]}
  \BibitemShut {NoStop}%
\bibitem [{\citenamefont {Duan}\ \emph {et~al.}(2019)\citenamefont {Duan} \emph
  {et~al.}}]{Duan2019}%
  \BibitemOpen
  \bibfield  {author} {\bibinfo {author} {\bibfnamefont {K.-K.}\ \bibnamefont
  {Duan}} \emph {et~al.},\ }\bibfield  {title} {\bibinfo {title} {{DmpIRFs and
  DmpST: DAMPE Instrument Response Functions and Science Tools for Gamma-Ray
  Data Analysis}},\ }\href {https://doi.org/10.1088/1674-4527/19/9/132}
  {\bibfield  {journal} {\bibinfo  {journal} {Res. Astron Astrophys.}\ }\textbf
  {\bibinfo {volume} {19}},\ \bibinfo {pages} {132} (\bibinfo {year} {2019})},\
  \Eprint {https://arxiv.org/abs/1904.13098} {arXiv:1904.13098 [astro-ph.HE]}
  \BibitemShut {NoStop}%
\bibitem [{\citenamefont {Zhang}\ \emph {et~al.}(2011)\citenamefont {Zhang}
  \emph {et~al.}}]{Acc2011}%
  \BibitemOpen
  \bibfield  {author} {\bibinfo {author} {\bibfnamefont {Y.-L.}\ \bibnamefont
  {Zhang}} \emph {et~al.},\ }\bibfield  {title} {\bibinfo {title} {Evaluation
  of particle acceptance for space particle telescope},\ }\href@noop {}
  {\bibfield  {journal} {\bibinfo  {journal} {CPC}\ }\textbf {\bibinfo {volume}
  {35}},\ \bibinfo {pages} {774} (\bibinfo {year} {2011})}\BibitemShut
  {NoStop}%
\bibitem [{\citenamefont {{Ackermann}}\ \emph {et~al.}(2013)\citenamefont
  {{Ackermann}}, \citenamefont {{Ajello}}, \citenamefont {{Albert}},
  \citenamefont {{Allafort}}, \citenamefont {{Baldini}} \emph
  {et~al.}}]{Ackermann2013a}%
  \BibitemOpen
  \bibfield  {author} {\bibinfo {author} {\bibfnamefont {M.}~\bibnamefont
  {{Ackermann}}}, \bibinfo {author} {\bibfnamefont {M.}~\bibnamefont
  {{Ajello}}}, \bibinfo {author} {\bibfnamefont {A.}~\bibnamefont {{Albert}}},
  \bibinfo {author} {\bibfnamefont {A.}~\bibnamefont {{Allafort}}}, \bibinfo
  {author} {\bibfnamefont {L.}~\bibnamefont {{Baldini}}}, \emph {et~al.}
  (\bibinfo {collaboration} {Fermi-LAT Collaboration}),\ }\bibfield  {title}
  {\bibinfo {title} {{Search for gamma-ray spectral lines with the Fermi Large
  Area Telescope and dark matter implications}},\ }\href
  {https://doi.org/10.1103/PhysRevD.88.082002} {\bibfield  {journal} {\bibinfo
  {journal} {Phys. Rev. D}\ }\textbf {\bibinfo {volume} {88}},\ \bibinfo {eid}
  {082002} (\bibinfo {year} {2013})},\ \Eprint
  {https://arxiv.org/abs/1305.5597} {arXiv:1305.5597} \BibitemShut {NoStop}%
\bibitem [{\citenamefont {{Albert}}\ \emph {et~al.}(2014)\citenamefont
  {{Albert}}, \citenamefont {{G{\'o}mez-Vargas}}, \citenamefont {{Grefe}},
  \citenamefont {{Mu{\~n}oz}} \emph {et~al.}}]{Albert2014}%
  \BibitemOpen
  \bibfield  {author} {\bibinfo {author} {\bibfnamefont {A.}~\bibnamefont
  {{Albert}}}, \bibinfo {author} {\bibfnamefont {G.~A.}\ \bibnamefont
  {{G{\'o}mez-Vargas}}}, \bibinfo {author} {\bibfnamefont {M.}~\bibnamefont
  {{Grefe}}}, \bibinfo {author} {\bibfnamefont {C.}~\bibnamefont
  {{Mu{\~n}oz}}}, \emph {et~al.},\ }\bibfield  {title} {\bibinfo {title}
  {{Search for 100 MeV to 10 GeV {\ensuremath{\gamma}}-ray lines in the
  Fermi-LAT data and implications for gravitino dark matter in the
  {\ensuremath{\mu}}{\ensuremath{\nu}}SSM}},\ }\href
  {https://doi.org/10.1088/1475-7516/2014/10/023} {\bibfield  {journal}
  {\bibinfo  {journal} {J. Cosmol. Astropart. Phys.}\ }\textbf {\bibinfo
  {volume} {10}}\bibfield  {number} {\bibinfo  {number} { (10)},\ \bibinfo
  {eid} {023}},\ }\Eprint {https://arxiv.org/abs/1406.3430} {arXiv:1406.3430}
  \BibitemShut {NoStop}%
\bibitem [{\citenamefont {Zhang}\ \emph {et~al.}(2019)\citenamefont {Zhang}
  \emph {et~al.}}]{Zhang2019}%
  \BibitemOpen
  \bibfield  {author} {\bibinfo {author} {\bibfnamefont {Y.-Q.}\ \bibnamefont
  {Zhang}} \emph {et~al.},\ }\bibfield  {title} {\bibinfo {title} {{Design and
  on-orbit status of the trigger system for the DAMPE mission}},\ }\href@noop
  {} {\bibfield  {journal} {\bibinfo  {journal} {Res. Astron Astrophys.}\
  }\textbf {\bibinfo {volume} {19}},\ \bibinfo {pages} {023} (\bibinfo {year}
  {2019})}\BibitemShut {NoStop}%
\bibitem [{\citenamefont {{Einasto}}(1965)}]{Einasto1965}%
  \BibitemOpen
  \bibfield  {author} {\bibinfo {author} {\bibfnamefont {J.}~\bibnamefont
  {{Einasto}}},\ }\bibfield  {title} {\bibinfo {title} {{On the Construction of
  a Composite Model for the Galaxy and on the Determination of the System of
  Galactic Parameters}},\ }\href@noop {} {\bibfield  {journal} {\bibinfo
  {journal} {Trudy Astrofizicheskogo Instituta Alma-Ata}\ }\textbf {\bibinfo
  {volume} {5}},\ \bibinfo {pages} {87} (\bibinfo {year} {1965})}\BibitemShut
  {NoStop}%
\bibitem [{\citenamefont {{Navarro}}\ \emph {et~al.}(2010)\citenamefont
  {{Navarro}}, \citenamefont {{Ludlow}}, \citenamefont {{Springel}},
  \citenamefont {{Wang}} \emph {et~al.}}]{Navarro2010}%
  \BibitemOpen
  \bibfield  {author} {\bibinfo {author} {\bibfnamefont {J.~F.}\ \bibnamefont
  {{Navarro}}}, \bibinfo {author} {\bibfnamefont {A.}~\bibnamefont {{Ludlow}}},
  \bibinfo {author} {\bibfnamefont {V.}~\bibnamefont {{Springel}}}, \bibinfo
  {author} {\bibfnamefont {J.}~\bibnamefont {{Wang}}}, \emph {et~al.},\
  }\bibfield  {title} {\bibinfo {title} {{The diversity and similarity of
  simulated cold dark matter haloes}},\ }\href
  {https://doi.org/10.1111/j.1365-2966.2009.15878.x} {\bibfield  {journal}
  {\bibinfo  {journal} {Mon. Not. R. Astron. Soc.}\ }\textbf {\bibinfo {volume}
  {402}},\ \bibinfo {pages} {21} (\bibinfo {year} {2010})},\ \Eprint
  {https://arxiv.org/abs/0810.1522} {arXiv:0810.1522} \BibitemShut {NoStop}%
\bibitem [{\citenamefont {{Acero}}\ \emph {et~al.}(2016)\citenamefont
  {{Acero}}, \citenamefont {{Ackermann}}, \citenamefont {{Ajello}},
  \citenamefont {{Albert}} \emph {et~al.}}]{FermiGDE2016}%
  \BibitemOpen
  \bibfield  {author} {\bibinfo {author} {\bibfnamefont {F.}~\bibnamefont
  {{Acero}}}, \bibinfo {author} {\bibfnamefont {M.}~\bibnamefont
  {{Ackermann}}}, \bibinfo {author} {\bibfnamefont {M.}~\bibnamefont
  {{Ajello}}}, \bibinfo {author} {\bibfnamefont {A.}~\bibnamefont {{Albert}}},
  \emph {et~al.} (\bibinfo {collaboration} {Fermi-LAT Collaboration}),\
  }\bibfield  {title} {\bibinfo {title} {{Development of the Model of Galactic
  Interstellar Emission for Standard Point-source Analysis of Fermi Large Area
  Telescope Data}},\ }\href {https://doi.org/10.3847/0067-0049/223/2/26}
  {\bibfield  {journal} {\bibinfo  {journal} {Astrophys. J. Suppl. Ser.}\
  }\textbf {\bibinfo {volume} {223}},\ \bibinfo {eid} {26} (\bibinfo {year}
  {2016})},\ \Eprint {https://arxiv.org/abs/1602.07246} {arXiv:1602.07246
  [astro-ph.HE]} \BibitemShut {NoStop}%
\bibitem [{\citenamefont {{Abdollahi}}\ \emph {et~al.}(2020)\citenamefont
  {{Abdollahi}}, \citenamefont {{Acero}}, \citenamefont {{Ackermann}},
  \citenamefont {{Ajello}}, \citenamefont {{Atwood}} \emph
  {et~al.}}]{4FGL2019}%
  \BibitemOpen
  \bibfield  {author} {\bibinfo {author} {\bibfnamefont {S.}~\bibnamefont
  {{Abdollahi}}}, \bibinfo {author} {\bibfnamefont {F.}~\bibnamefont
  {{Acero}}}, \bibinfo {author} {\bibfnamefont {M.}~\bibnamefont
  {{Ackermann}}}, \bibinfo {author} {\bibfnamefont {M.}~\bibnamefont
  {{Ajello}}}, \bibinfo {author} {\bibfnamefont {W.~B.}\ \bibnamefont
  {{Atwood}}}, \emph {et~al.} (\bibinfo {collaboration} {Fermi-LAT
  Collaboration}),\ }\bibfield  {title} {\bibinfo {title} {{Fermi Large Area
  Telescope Fourth Source Catalog}},\ }\href
  {https://doi.org/10.3847/1538-4365/ab6bcb} {\bibfield  {journal} {\bibinfo
  {journal} {Astrophys. J. Suppl. Ser.}\ }\textbf {\bibinfo {volume} {247}},\
  \bibinfo {eid} {33} (\bibinfo {year} {2020})},\ \Eprint
  {https://arxiv.org/abs/1902.10045} {arXiv:1902.10045 [astro-ph.HE]}
  \BibitemShut {NoStop}%
\end{thebibliography}%

\end{document}